\documentclass[11pt,twoside]{article}
\usepackage{./asp2014}

\aspSuppressVolSlug
\resetcounters

\begin{document}

\title{Significantly super-Chandrasekhar limiting mass white dwarfs and 
their consequences}
\author{B. Mukhopadhyay,$^1$ U. Das,$^2$ A. R. Rao,$^3$ S. Subramanian,$^4$ 
M. Bhattacharya,$^5$ S. Mukerjee,$^1$ T.S. Bhatia,$^1$ and J. Sutradhar$^1$
\affil{$^1$Department of Physics, Indian Institute of Science, Bangalore, India; 
\email{bm@physics.iisc.ernet.in
}}
\affil{$^2$JILA, University of Colorado, Boulder, USA; \email{upasana.das@jila.colorado.edu}}
\affil{$^3$Tata Institute of Fundamental Research, Mumbai, India; \email{arrao@tifr.res.in}}
\affil{$^4$University of Cambridge, Cambridge, UK; \email{ss2310@cam.ac.uk}}
\affil{$^5$University of Texas, Austin, USA; \email{mukul.b@utexas.edu}}}

\paperauthor{B. Mukhopadhyay}{bm@physics.iisc.ernet.in}{ORCID_Or_Blank}{Indian Institute of Science}{Department of Physics}{Bangalore}{Karnataka}{560012}{India}
\paperauthor{U. Das}{upasana.das@jila.colorado.edu}{ORCID_Or_Blank}{University of Colorad}{JILA}{Boulder}{Colorado}{Postal Code}{USA}
\paperauthor{A. R. Rao}{arrao@tifr.res.in}{ORCID_Or_Blank}{Tata Institute of Fundamental Research}{Department of Astrophysics and Astronomy}{Mumbai}{Maharashtra}{40005}{India}
\paperauthor{S. Subramanian}{ss2310@cam.ac.uk}{ORCID_Or_Blank}{University of Cambridge}{Department}{Cambridge}{}{Postal Code}{UK}
\paperauthor{M. Bhattacharya}{mukul.b@utexas.edu}{ORCID_Or_Blank}{University of Texas}{Department}{Austin}{}{Postal Code}{USA}
\paperauthor{S. Mukherjee}{smukerjee@physics.iisc.ernet.in}{ORCID_Or_Blank}{Indian Institute of Science}{Department of Physics}{Bangalore}{Karnataka}{560012}{India}
\paperauthor{T. S. Bhatia}{tanayveer1@gmail.com}{ORCID_Or_Blank}{Indian Institute of Science}{Department of Physics}{Bangalore}{Karnataka}{560012}{India}
\paperauthor{J. Sutradhar}{jagu@physics.iisc.ernet.in}{ORCID_Or_Blank}{Indian Institute of Science}{Department of Physics}{Bangalore}{Karnataka}{560012}{India}

\begin{abstract}

Since 2012, we have initiated a new idea
showing that the mass of highly magnetized 
or modified Einstein's gravity induced white dwarfs could be significantly 
super-Chandrasekhar with a different mass-limit. This discovery has several important 
consequences, including explanation of peculiar, over-luminous 
type Ia supernovae, soft gamma-ray repeaters and anomalous X-ray pulsars without 
invoking extraordinarily strong, yet unobserved, magnetic fields. 
It further argues for a possible second standard candle. 
Based on simpler calculations, these white dwarfs are also shown to be much 
less luminous than their standard counter-parts 
(of low magnetic fields). 
This discovery altogether initiates a new field of research.

\end{abstract}

\section{Introduction}

Since 2012, we have initiated exploring highly magnetized super-Chandrasekhar white dwarfs 
(B-WDs). The primary aim behind this was explaining peculiar, over-luminous type Ia 
supernovae (SNeIa). However, subsequently they were found to be useful to explain 
other data, e.g. soft gamma-ray repeaters (SGRs) and anomalous X-ray pulsars (AXPs),
white dwarf (WD) pulsar(s), etc.
This immediately brings the topic super-Chandrasekhar WDs in lime-light,
with so many groups' coming forward to work in this new field, who 
need not be focusing on magnetic effects only (just to mention
a very few out of the bulk, \citealt{liu,schr,emil}). 
In order to establish this field,
our approach has been, so far, the following.

First, we have considered most simplistic, spherically symmetric, 
highly magnetized B-WDs in the Newtonian framework. 
This brings quantum mechanical effects
in the equation of state (EoS), in the presence of high amplitude of field (\citealt{prd,prl}). 
In the same model, we have also shown that B-WDs
altogether have a new mass-limit, $80\%$ larger than the Chandrasekhar-limit, 
in the same spirit as the Chandrasekhar-limit was obtained (\citealt{chandra}). 
Afterwards, we removed the assumptions of Newtonian description
and spherical symmetry (e.g. \citealt{jcap1,jcap2}). 
Based on a full scale general relativistic magnetohydrodynamic (GRMHD) description 
(\citealt{jcap2,mnras}), 
we explored more self-consistent B-WDs that are ellipsoids and have revealed 
similar masses as obtained in the simpler framework.

In a different avenue, we also explored modified Einstein's gravity 
(Starobinsky model) induced model (\citealt{jcap3}) to unify under- and over-luminous 
SNeIa. We would not touch this work in the limited scope of this 
proceedings.

To follow the motivation of this work and systematic progress of the topic 
in detail, the readers are advised to see \citealt{colo}, and the references therein (also see, \citealt{ost}). 
Here we touch upon the basic results.

\section{Effect of magnetic field via equation of state}

We assume the magnetic field to be fluctuating in such a way that effective field
brings negligible effect over matter pressure. Hence, they do not
contribute to MHD. However, the fluctuating length scale is larger than the 
Compton wavelength of electrons ($\lambda_{e}$) so that Landau quantization affects the 
degenerate electron gas EoS, given by
\begin{eqnarray}
&&P =  \frac{2B_D}{(2\pi)^{2}\lambda_{e}^{3}}m_{e}c^{2}\sum_{\nu=0}^{\nu_{m}} g_{\nu} (1 + 2\nu B_{D}) \eta\left(\frac{x_{F}(\nu)}{(1 + 2\nu B_{D})^
{1/2}} \right), \\
\label{pressure}
&&{\eta}(y) = \frac{1}{2}y\sqrt{1 + y^{2}} - \frac{1}{2}\ln(y+ \sqrt{1 + y^{2}}),
\nonumber
\end{eqnarray}
where $B_D=B/4.414\times 10^{13}$G. Other variables have their usual meanings, see
\citealt{colo}, for details. At high density (e.g. around the center of star), EoS
approximately reduces to $P=K_m(B)\rho^2$.

The underlying magnetized, spherical B-WD obeys the magnetostatic equilibrium
condition, along with estimate of mass, as
\begin{eqnarray}
\frac{1}{\rho+\rho_B}\frac{d}{dr}\left(P+\frac{B^2}{8\pi}\right)=F_g+\left.\frac{\vec{B}\cdot\nabla\vec{B}}{4\pi(\rho+\rho_B)}\right|_r,\,\,\,\,
\label{magstat}
\frac{dM}{dr}=4\pi r^2(\rho+\rho_B).
\label{mas}
\end{eqnarray}
For the present purpose, magnetic terms could be neglected in above equations
and following Lane-Emden formalism (\citealt{prl}),
the scalings of mass and radius
with central density $\rho_c$ are obtained as
\begin{eqnarray}
M \propto K_m^{3/2} \rho_c^{(3-n)/2n},\,\,\,\, R\propto K_m^{1/2}\rho_c^{(1-n)/2n},
\,\,\,\, K_m=K\rho_c^{-2/3}. 
\label{scal}
\end{eqnarray}
Clearly $n=1$ ($\Gamma=2$) corresponds to $\rho_c$-independent $M$ (unlike 
the Chandrasekhar's case when $K_m$ is independent of $B$ and limiting mass
corresponds to $n=3$).

Substituting proportionality constants appropriately,
we obtain the limiting mass
\begin{equation}
M_l=\left(\frac{hc}{2G}\right)^{3/2}\frac{1}{(\mu_e m_H)^2}
\approx \frac{10.312}{\mu_e^2}M_\odot,
\label{mass3}
\end{equation}
when the limiting radius $R_l\rightarrow 0$.
For $\mu_e=2$, $M_l=2.58M_\odot$. Importantly,
for finite but high density and magnetic field, e.g. $\rho_c=2\times 10^{10}$gm/cc
and $B=8.8\times 10^{15}$G when $E_{Fmax}=20m_ec^2$, $M=2.44M_\odot$ and $R$ is about 650km. Note that
these $\rho_c$ and $B$ are below their respective upper limits set by the instabilities of
pycnonuclear fusion, inverse-$\beta$ decay and general relativistic 
effects (\citealt{mpla}). 

\section{Effect of magnetic field via MHD}

\begin{figure}[h]
   \centering
\includegraphics[angle=0,scale=0.33]{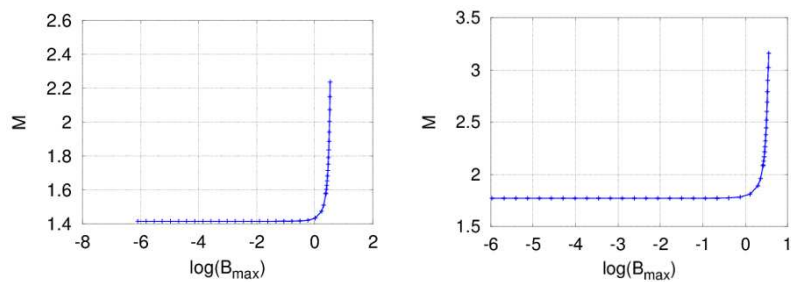}
\caption{
Non-rotating (left) and rotating (right) sequences of mass in $M_\odot$
with changing maximum field in G, for toroidal magnetic fields. Chosen 
central angular velocity is $30.42$ rad/sec. See \citealt{mnras}, for details.
}
\label{seq}
\end{figure}

When magnetic field is not fluctuating, magnetic pressure and density cannot be
neglected in the stellar structure equations. 
We model such B-WDs using the publicly available 
{\it XNS} code (\citealt{pili}).
In Fig. \ref{seq} we present some representative results, reported in detail earlier 
(\citealt{jcap2,mnras,colo}), showing again that
B-WDs could be significantly super-Chandrasekhar. Note however that we restrict the central 
field in such a way that the ratio of magnetic to gravitational energies is significantly
below unity. This restriction furthermore hinders EoS to be modified by Landau
quantization. 

\section{Luminosity of B-WDs}

Now following an established technique (\citealt{st}), we divide B-WDs into 
inner core having degenrate EoS (for the present purpose non-relativistic)
and outer envelope having ideal gas EoS. 
We then solve the magnetostatic equilibrium and photon diffusion equations 
in the presence of a magnetic field given by
\begin{eqnarray}
&&\frac{d}{dr}(P+P_{B}) = -\frac{GM}{r^2}(\rho+\rho_{B})\,\,\,
{\rm and}\,\,\,
\frac{dT}{dr} = -\frac{3}{4ac}\frac{\kappa_{0}(\rho+\rho_{B})^2}{T^{6.5}}\frac{L}{4\pi r^2},
\\
\nonumber
&&{\rm with}\,\,\,
B\left(\frac{\rho}{\rho_{0}}\right) = B_{s} + B_{0}\left[1-\rm{exp}{\left(-\eta \left(\frac{\rho}{\rho_{0}}\right)^{\gamma}\right)}\right],
\,\,
\rho_{*} \approx (2.4\times10^{-8} {\rm{\:g\:cm^{-3}}})\: \mu_{e}T_{*}^{3/2},
\label{bale}
\end{eqnarray}
and investigate 
the temperature profile in the envelope. 
Here $B_{s}$ is the surface magnetic field and $B_0$ is similar to the central field $B_c$, 
the value of $\rho_{0}$ is chosen to be $10\%$ of $\rho_{c}$, 
also $\rho_*$ and $T_*$ are the interface density and temperature respectively.
We also fix surface density 
$\rho_s=10^{-9}$ gm/cc and radius $R=5000$ km and solve above equations simultaneously,
for non-magnetic WDs and B-WDs separately. Interestingly,
for fixed interface radius and/or temperature between magnetic and non-magnetic
cases, B-WDs turn out to be much less luminous, as seen in Table 1. This is roughly
understood, in this simplistic model, from the magnetostatic balance condition for similar mass and radius
between two cases. Generally for $B$ under consideration, $\rho_B<<\rho$ but 
$P\sim \rho_B$. Hence, for a similar gravitational field, B-WDs have to have a
smaller thermal energy and hence luminosity ($L$). As $L<10^{-5}L_\odot$
cannot be detected yet, B-WDs with $L\lesssim 10^{-6}L_\odot$, as given in Table 1,
appear to be invisible. 

\begin{table*}[htbp]
\begin{center}
\caption{\small Variation of luminosity with magnetic field for a fixed interface radius.
For other details, see \citealt{mukul}.}
\begin{tabular}{cccccccccccccccccccccc}
\hline
\hline
\centering
$B=(B_s,B_c)\:({\rm{in\:G}})$ & $L\:(\rm{in}$$\:L_{\odot})$ &  $T_{*}\:\rm{(in\:K)}$ & $\rho_{*}\:\rm{(in\:g\:cm^{-3})}$ & $T_{s}\:\rm{(in\:K)}$ \\ \hline
$B=(0, 0)$                 & $10^{-5}$ & $2.332\times10^{6}$ & $170.722$ & $3850$ \\ \hline
$B=(10^{9}, 10^{13})\: $ & $5.17\times10^{-6}$ & $1.94346\times10^{6}$ & $129.886$ & $3260$\\ \hline
$B=(10^{9}, 5\times10^{13})\: $ & $2.87\times10^{-8}$ & $495107$ & $16.7012$ & $890$\\ \hline
$B=(10^{10}, 5\times10^{12})\: $ & $1.35\times10^{-6}$ & $1.37451\times10^{6}$ & $77.2534$ & $2330$\\ \hline
$B=(10^{11}, 10^{13})\: $ & $2.33\times10^{-13}$ & $70750.2$ & $0.902173$ & $50$\\ \hline
\hline
\end{tabular}
\end{center}
\end{table*}

\section{B-WDs as SGRs/AXPs}

SGRs/AXPs are polularly explained by magnetar model (\citealt{thompson}).
However, there are many shortcomings in the magnetar model, see \citealt{mareg},
for details. Weakly magnetized WD based model is challenged by 
observed short spin periods and low UV-luminosities ($L_{UV}$). 
We explore the possibility to explain the high energy phenomena in AXPs/SGRs
by rotationally powered magnetic energy ($\dot{E}_{dot}$) of B-WDs --
there is no need to invoke extraordinary, 
yet observationally unconfirmed, sources of energy. 
This is possible because B-WDs 
have larger moment of inertia than neutron stars, 
which is however small enough to produce $L_{UV}$.

\begin{figure}[h]
\vskip 0.3in
   \centering
\includegraphics[angle=0,scale=0.4]{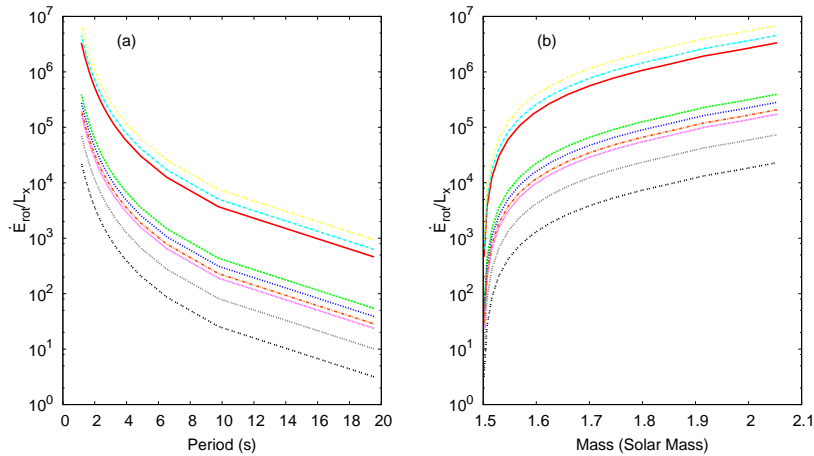}
\vskip0.5cm
\caption{The ratio of rate of rotational energy release to X-ray luminosity as a function of
(a) spin period, and (b) mass, for B-WDs
when from the top to bottom various curves correspond to 1E 1547-54, 1E 1048-59,
SGR 1806-20, SGR 1900+14, SGR 0526-66, SGR 1822-1606, 1E 1841-045, SGR 0418+5729 and 1E 2259+586.
For other details, see \citealt{jcap4}.
}
\label{hmedot}
\end{figure}
Figure \ref{hmedot} shows that $\dot{E}_{\rm rot}$ computed based on B-WD model, 
with a fixed inclination angle between rotation and magnetic axes $\alpha=15^\circ$, 
is several orders of magnitude larger than observed X-ray luminosity 
$L_x$ for nine sources. 

\section{Are GCRT J1745-3009 and AR Sco B-WDs?}

The transient radio source GCRT J1745-3009 was argued earlier to be a WD pulsar 
(\citealt{zhang}), but this idea was ruled out by color-magnitude analysis 
(\citealt{kaplan}). Now in the framework of a very slowly rotating B-WD
with $B_s\sim 3.3\times 10^{11} - 2\times 10^{12}$ G,
corresponding $R\sim 1580 - 500$ km 
and $\rho_c\sim 10^{10}$ gm/cc, we revisit all
the calculations, e.g. radius of polar cap and unipolar potential drop therein, etc.,
done by \cite{zhang}, and find them to be consistent with 
WD pulsar idea (see \citealt{jcap4}, for details). 
The maximum gamma-ray/X-ray
flux appears to be only a factor of 4 larger than that obtained earlier (\citealt{zhang}) 
for the same parameters for very highly magnetized B-WDs. 
The condition of radio luminosity not exceeding the spin-down luminosity reveals
the distance of the source to be $\lesssim 8.5$ kpc, 
which is in accordance with the lower limit predicted previously (\citealt{kaplan}). 
As shown in \S 4,
such a B-WD is significantly cooler, 
hence its optical flux will be dimmer to evade detection, 
strongly supporting it to be a WD pulsar.

Recently, AR Sco has been found to be a WD pulsar with spin period 1.95 min 
(\citealt{nature}).
While a WD following Chandrasekhar's mass-radius relation with mass $0.8-1.29M_\odot$
could explain the source, in order to explain its emission as spin-down power,
its $B_s$ could be $\sim 10^9$ G and hence $B_c$ could be $10^{12}$ G. 
However, if eventually accretion process starts in AR Sco (e.g. by WD's coming closer
to the companion by the emission of gravitational radiation),
based on our past result (\citealt{rao}), initial smaller $(B_s,B_c)$ may enhance 
via flux-freezing and WD may deviate from Chandrasekhar's mass-radius 
path leading to a B-WD. Hence, AR Sco is plausibly a seed of B-WD.

\section{Critique of B-WDs}

Since the birth of this field, while several groups have been supporting
it with follow-up work (e.g. \citealt{liu,schr,emil}, to mention a very
few), there are some critics as well. While some criticism
(\citealt{konar}) was found to be based on erroneous calculations and was
dismissed immediately (\citealt{reply}), some others
(\citealt{chamel,ruffini}) were shown to be misleading
(\citealt{mpla,colo}). Recently, a new issue has been brought
up (\citealt{bera}), arguing that such magnetized WDs are
unstable. However, this argument is based on purely poloidal (or highly poloidally
dominated)
and purely toroidal field configurations, which have long been proposed to
be unstable (\citealt{tayler}, also see the comments in \citealt{jcap2}).
They are not expected to be naturally occurring configurations. 
On a technical note, some of the non-linear
perturbation analysis carried out by \cite{bera} seems to show the non-perturbed
equilibrium state itself to be evolving with time, which also raises a question 
regarding the robustness of the method employed.
Moreover, B-WDs
could have fluctuating fields and, hence, overall low average fields, as discussed
in section
2, which do not encounter such a problem, if any.
We believe that one should look for plausible mixed field configurations
(\citealt{ciolfi}) with self-consistent inclusion of rotation, to perform
such a stability analysis, before drawing any bold conclusions.

\section{Conclusion}
   
We have initiated a new field establishing the possible formation of
highly super-Chandrasekhar magnetized WDs, which will also have a
new mass-limit. Such WDs have several important implications: formation of
peculiar overluminous SNeIa, SGRs/AXPs, WD pulsars, to name a few. Our
future aim is to systematically unfold all the issues related to such WDs, including
self-consistent stability analysis based on realistic magnetic field geometries, exploring
their actual connection to over-luminous SNeIa etc. Hence, we welcome the
community to join us to work in this fascinating new field.



\begin{thebibliography}{}
\bibitem[Belyaev et al. (2015)]{emil} Belyaev, V. B., Ricci, P., Simkovic, F., Adam, J., 
Tater, M., \& Truhlik, E. 2015, Nuc. Phys. A, 937, 17.
\bibitem[Bera \& Bhattacharya (2016)]{bera} Bera, P., \& Bhattacharya, D. 2016;
arXiv:1607.06829v1.
\bibitem[Bhattacharya et al. (2015)]{mukul} Bhattacharya, M., Mukhopadhyay, B., \&
Mukerjee, S. 2015; arXiv:1509.00936.
\bibitem[Chamel et al. (2013)]{chamel} Chamel, N., Fantina, A.F., \& Davis, P.J. 2013,
Phys. Rev. D, 88, 081301.
\bibitem[Chandrasekhar (1935)]{chandra} Chandrasekhar, S. 1935,
MNRAS, 95, 207.
\bibitem[Ciolfi \& Rezzolla (2013)]{ciolfi} Ciolfi, R., \& Rezzolla, L. 2013, MNRAS, 435, L43.
\bibitem[Das \& Mukhopadhyay (2012)]{prd} Das, U., \& Mukhopadhyay, B. 2012,
Phys. Rev. D, 86, 042001.
\bibitem[Das \& Mukhopadhyay (2013)]{prl} Das, U., \& Mukhopadhyay, B. 2013,
Phys. Rev. Lett., 110, 071102.
\bibitem[Das \& Mukhopadhyay (2014a)]{jcap1} Das, U., \& Mukhopadhyay, B. 2014a,
JCAP, 06, 050. 
\bibitem[Das \& Mukhopadhyay (2014b)]{mpla} Das, U., \& Mukhopadhyay, B. 2014b, 
MPLA, 29, 1450035.
\bibitem[Das \& Mukhopadhyay (2015a)]{jcap2} Das, U., \& Mukhopadhyay, B. 2015a,
JCAP, 05, 016. 
\bibitem[Das \& Mukhopadhyay (2015b)]{jcap3} Das, U., \& Mukhopadhyay, B. 2015b,
JCAP, 05, 045. 
\bibitem[Das \& Mukhopadhyay (2015c)]{reply} Das, U., \& Mukhopadhyay, B. 2015c,
Phys. Rev. D, 91, 028302.
\bibitem[Das et al. (2013)]{rao} Das, U., Mukhopadhyay, B., \& Rao, A.R. 2013,
ApJ, 767, L14.
\bibitem[Duncan \& Thompson (1992)]{thompson} Duncan, R.C., \& Thompson, C. 1992, 
ApJ, 392, L9.
\bibitem[Franzon \& Schramm (2015)]{schr}
Franzon, B., \& Schramm, S. 2015, Phys. Rev. D, 92, 083006.
\bibitem[Kaplan et al. (2008)]{kaplan} Kaplan, D.L., et al. 2008, ApJ, 687, 262.
\bibitem[Liu et al. (2014)]{liu}
Liu, H., Zhang, X., \& Wen, D. 2014, Phys. Rev. D, 89, 104043. 
\bibitem[Maleiro et al. (2012)]{ruffini} Malheiro, M., Rueda, J.A., \& Ruffini, R. 2012, PASJ, 64, 56.
\bibitem[Marsh et al. (2016)]{nature}
Marsh, T.R., et al. 2016, Nature, 537, 374.
\bibitem[Mereghetti (2012)]{mareg} Mereghetti, S. 2012, in Proceedings of the 26th Texas Symposium on Relativistic Astrophysics,
Sao Paulo, December 16-20, 2012; arXiv1304.4825.
\bibitem[Mukhopadhyay (2015)]{colo} Mukhopadhyay, B. 
2015, in Proc. CIPANP2015, Vail, CO, USA, May 19-24, 2015; arXiv:1509.09008.
\bibitem[Mukhopadhyay \& Rao (2016)]{jcap4} Mukhopadhyay, B., \& Rao, A.R. 2016,
JCAP, 05, 007. 
\bibitem[Nityananda \& Konar (2014)]{konar} Nityananda, R., \& Konar, S. 2014,
Phys. Rev. D, 89, 103017.
\bibitem[Ostriker \& Hartwick (1968)]{ost} Ostriker, J.P., \& 
Hartwick, F.D.A. 1968, ApJ, 153, 797.
\bibitem[Pili et al. (2014)]{pili} Pili, A.G., Bucciantini, N., Del Zanna, L.
2014, MNRAS, 439, 3541.
\bibitem[Shapiro \& Teukolsky (1983)]{st} Shapiro, S.L., \& Teukolsky, S.A. 1983, 
Black Holes, White Dwarfs, and Neutron Stars: The Physics of Compact Objects, Wiley, New York.
\bibitem[Subramanian \& Mukhopadhyay (2015)]{mnras} Subramanian, S., \& Mukhopadhyay, B. 
2015, MNRAS, 454, 752.
\bibitem[Tayler (1973)]{tayler} Tayler, R.J. 1973, MNRAS, 161, 365.
\bibitem[Zhang \& Gil (2005)]{zhang} Zhang, B., \& Gil, J. 2005, ApJ, 631, L143.

\end{thebibliography}


\end{document}